\documentclass[a4paper,11pt]{article}
\pdfoutput=1 

\usepackage{jheppub} 

\usepackage[T1]{fontenc} 
\usepackage{float}
\usepackage{ragged2e}

\title{SDHCAL technological prototype test beam results\\ \vspace{10pt} \small Talk presented at the International Workshop on Future Linear Colliders (LCWS2021), 15-18 March 2021. C21-03-15.1.}

\author[\dagger, 1]{H. García Cabrera\note{Speaker}. On behalf of the CALICE Collaboration.}

\affiliation[\dagger]{Centro de Investigaciones Energéticas Medioambientales y Tecnológicas (CIEMAT). Madrid, Spain.}

\emailAdd{hector.garcia2@ciemat.es}

\abstract{The Semi-Digital Hadronic Calorimeter (SDHCAL) is proposed to equip the future ILC detector. A technological prototype of the SDHCAL developed within the CALICE collaboration has been extensively tested in test beams. We review here the prototype performances in terms of hadronic shower reconstruction from the most recent analyses test beam data.}

\begin{document} 
\maketitle

\section{Introduction}

The CALICE SDHCAL technological prototype was developed compatible with the future \emph{International Linear Collider} (ILC) requirements and the \emph{Particle Flow Algorithm} (PFA) in mind. Its high granularity provides excellent tracking tools for PFA while keeping the requirements of efficiency, compactness and power consumption. During the last few years, several campaigns of data taking in test beam conditions without a magnetic field have been completed, resulting in analysis of the detector capabilities, energy resolution, corrections and particle identification algorithms. In the following, after a short description of the prototype, the test beam results will be presented.\\

\subsection{Prototype description}

\begin{figure}[t]
    \centering
    \includegraphics[width=\linewidth]{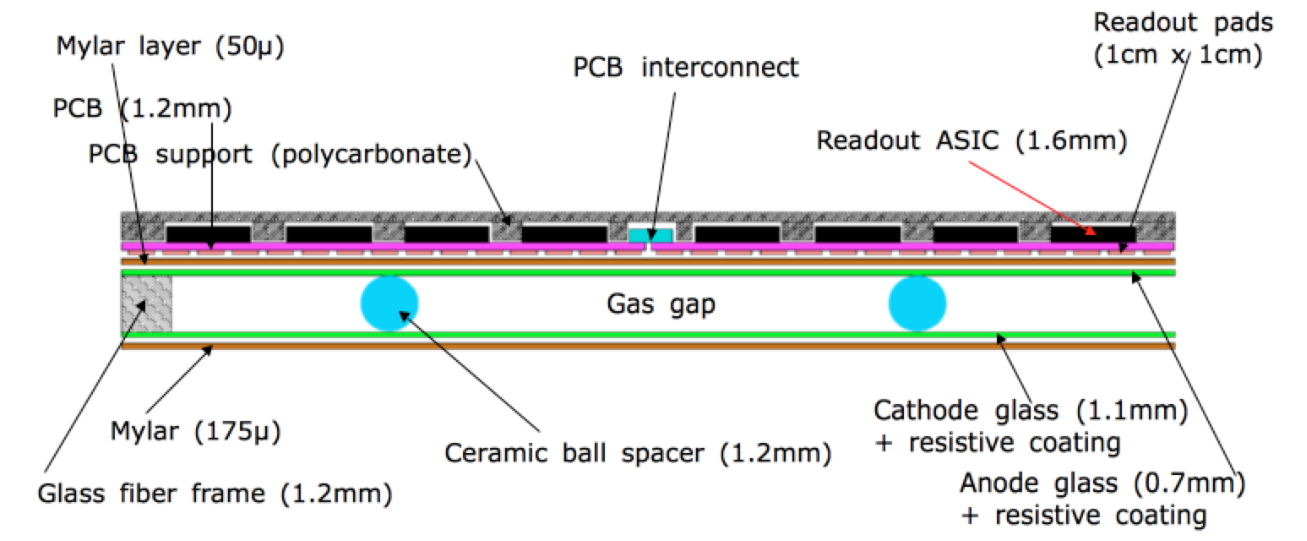}
    \caption{Schematics of a GRPC chamber with the readout electronics.}
    \label{fig:GRPC}
\end{figure}

The SDHCAL \cite{baulieu2015construction} is an hadronic calorimeter comprised of 48 active layers. Each layer is made of a 1 x 1 $m^2$ \emph{Glass Resistive Plate Chambers} (GRPC). The charge produced by ionization from incoming particles is read by a total of 9216 1 x 1 $cm^2$ silicon pads placed on the back of the layer, in front of the readout electronics as shown in Figure \ref{fig:GRPC}. The readout chip, HAdronic Rpc Detector ReadOut Chip (HARDROC) ASIC \cite{baulieu2015construction}, has a three analog thresholds that produce a digital 2 bit output. The active layer (GRPC) together with the readout electronics are placed inside a 2.5 mm stainless steel cassette which protects and facilitates handling of the layer while being part of the absorber. The cassettes are installed in a self-supporting mechanical structure built using 1.5 cm stainless steel plates that amounts to about 0.12 interaction lengths ($\lambda_I$) or 1.14 radiation lengths ($X_0$) per layer.\\

\section{Particle identification}
\label{sec:ParticleIdentification}

The SDHCAL prototype was exposed to pion, muon and electron beams in several data taking campaigns. The calorimeter worked in triggerless power-pulsing mode, in which the detector recorded all fired hits from the beginning until the end of a spill or up to the point were an ASIC's memory is filled. The resulting data sets are a combination of beam particles, cosmic rays contamination and electronic noise. To separate the different event types an analysis of the topology of the registered hits is performed \cite{2016}.\\

\noindent After the time reconstruction step \cite{baulieu2015construction, 2016} only the physically meaningful events remain. For this events the following selection variables are computed: 

\begin{itemize}
    \item \textbf{Density:} $\rho = \frac{N_{Hit}}{N_{Layers}}$ \hspace{5pt} --- \hspace{5pt} Where $N_{Layers}$ is only the layers with registered hits.
    \item \textbf{Second maximum of hits in a single layer:} $N_{Max2}$
    \item \textbf{Penetrability condition (P.C.):} True if the following is fulfilled $\rightarrow$ 
    \begin{itemize}
        \item[$\rhd$] Layers 01 - 10: At least 7 layers with signal.
        \item[$\rhd$] Layers 11 - 20: At least 7 layers with signal.
        \item[$\rhd$] Layers 21 - 35: At least 9 layers with signal.
        \item[$\rhd$] Layers 35 - 48: At least 8 layers with signal.
    \end{itemize}
    \item \textbf{Incident angle:} computed from the reconstructed track in the first 15 layers.
    \item \textbf{Shower start layer.}
\end{itemize}

\begin{figure}[h]
    \includegraphics[width=0.5\linewidth]{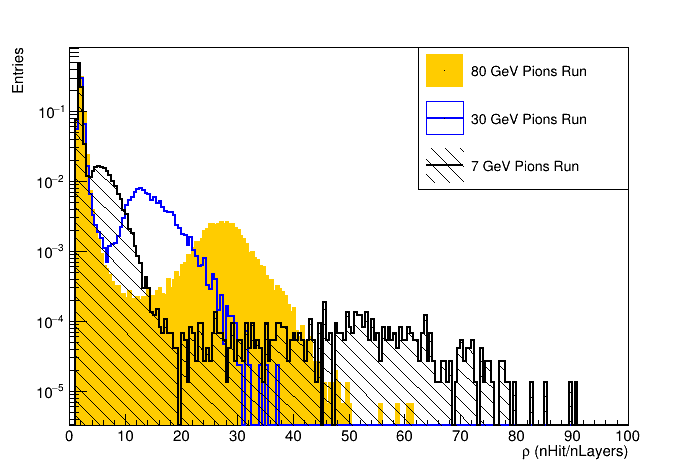}
    \includegraphics[width=0.5\linewidth]{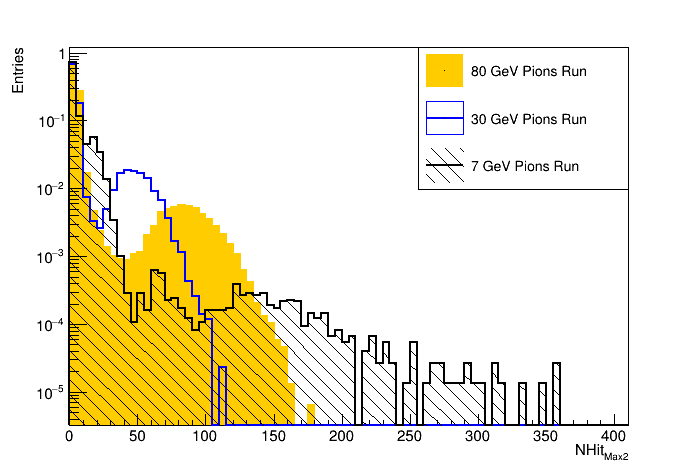}
    \caption{Density (left) and Second Maximum (right) variables for three different energy pion runs. These variables are powerful tools to separate the showers from cosmic rays and beam muons.}
    \label{fig:DensitySecondMax}
\end{figure}

\noindent Figure \ref{fig:DensitySecondMax} shows the density and second maximum in three pion runs of different energy. In all runs there are two peaks, the first one from beam muons and cosmic rays and the other appearing from particles showering inside the detector. By applying cuts to those variables it is possible to separate this topologies and together with the P.C. beam muons are identified from the cosmic rays.\\

\begin{figure}[t]
    \hspace{35pt}
    \includegraphics[width=0.4\linewidth]{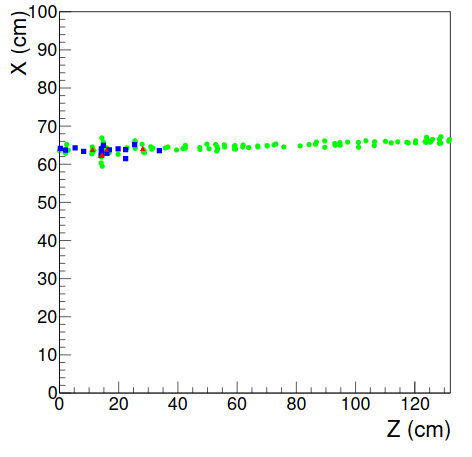}
    \includegraphics[width=0.385\linewidth]{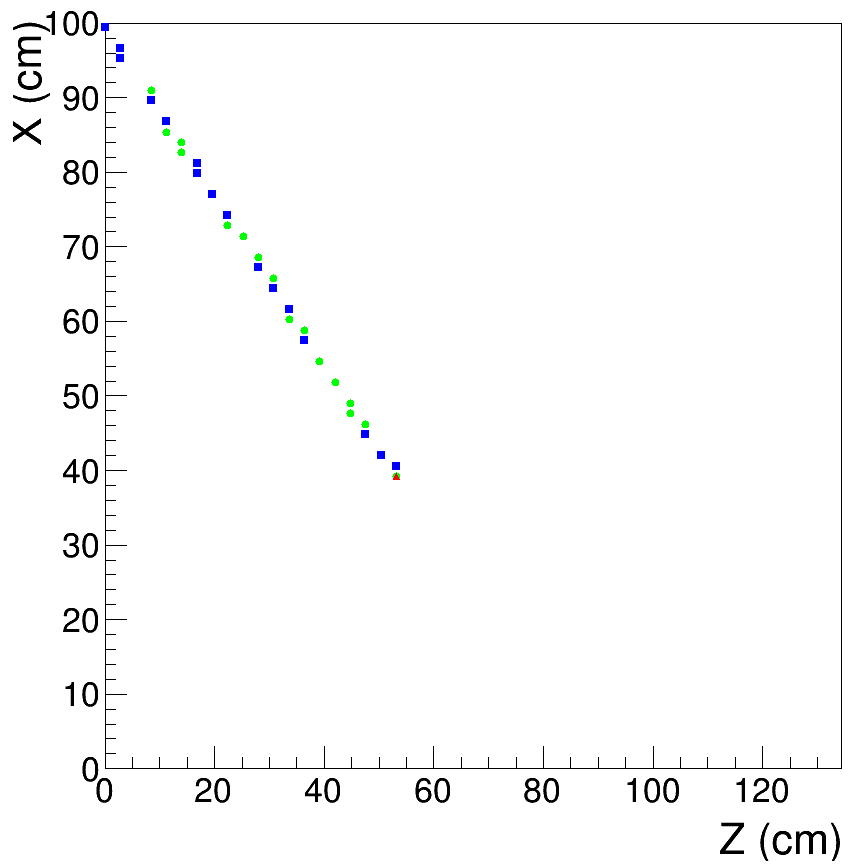}\\
    
    \hspace{35pt}
    \includegraphics[width=0.4\linewidth]{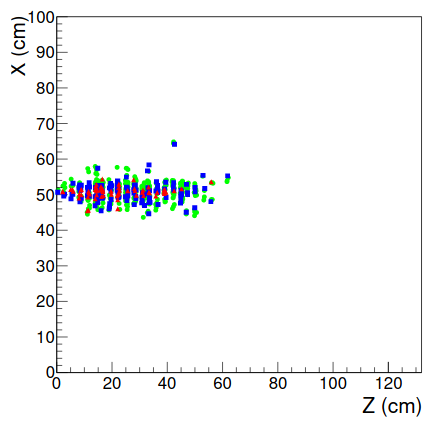}
    \includegraphics[width=0.4\linewidth]{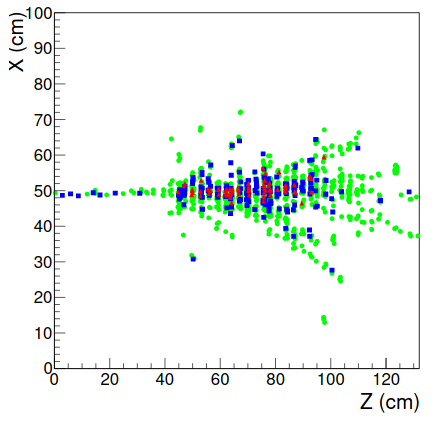}
    \caption{Different topologies identified in the detector by applying the selection cuts.}
    \label{fig:Toplogies}
\end{figure}

\noindent Now the events selected as showers are a mixture of remaining high density cosmic rays, pion showers and Electron showers. The first type is selected through the incident angle since cosmic rays enter the detector isotropically from all directions. Finally, since the interaction length of Electrons is much shorter than pions, if the shower start layer is in the fourth layer or beyond such event is identified as a pion shower. Figure \ref{fig:Toplogies} shows the event displays for this different toplogies selected once the selection cuts are applied to the data.\\

\section{Corrections and Performance}
\label{sec:CorrectionsPerformance}

From the previous section it is possible to identify specific particles in the data and then use its properties to analyze different effects in the detector, to correct them, and also to study the detectors performance. This section describes the different analysis, the results and consequential corrections applied to the data.

\subsection{Beam intensity correction}

\begin{figure}[b]
    \includegraphics[width=\linewidth]{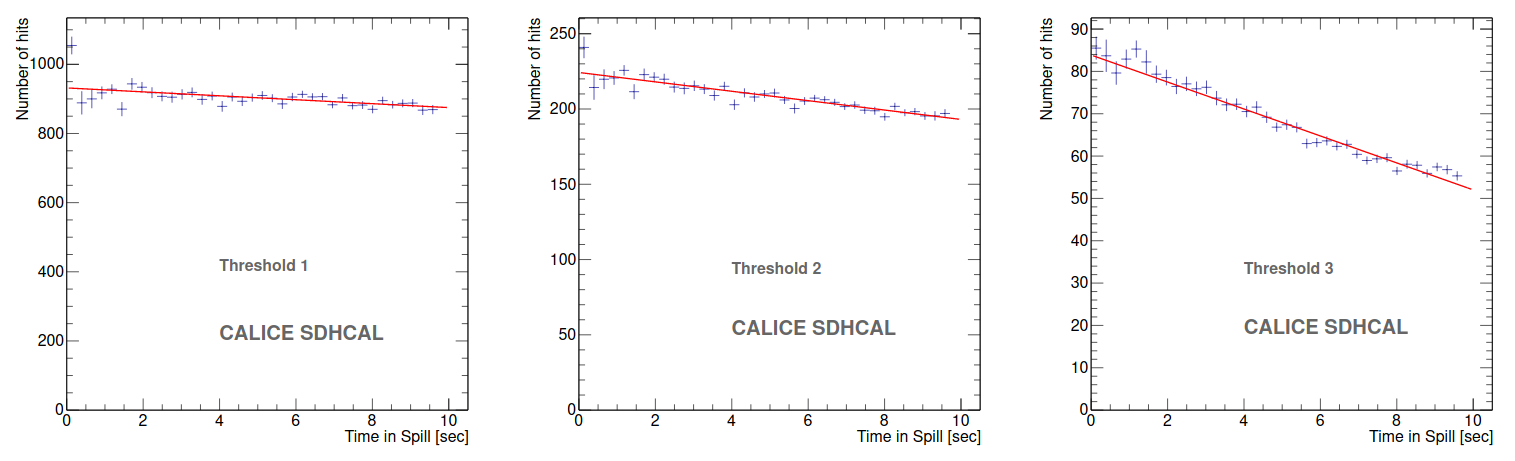}
    \caption{Beam intensity effect for each threshold in a 80 GeV pion run. Each time bin has the mean number of hits from all the events in the run.}
    \label{fig:Saturation}
\end{figure}

The beam parameters during the data taking campaigns were optimized to get spills with low particle rates. However, it was observed that the number of hits associated to hadronic showers were decreasing during the spill time since the particle rate increases during the spill time due to the structure of the spill \cite{2016}. The consequence of this effect is a degradation of the energy resolution in the showers. Figure \ref{fig:Saturation} displays this effect for each threshold, the decrease is more prominent for the number of hits associated to the second and third thresholds. \\

\begin{figure}[h]
\centering
    \includegraphics[width=0.75\linewidth]{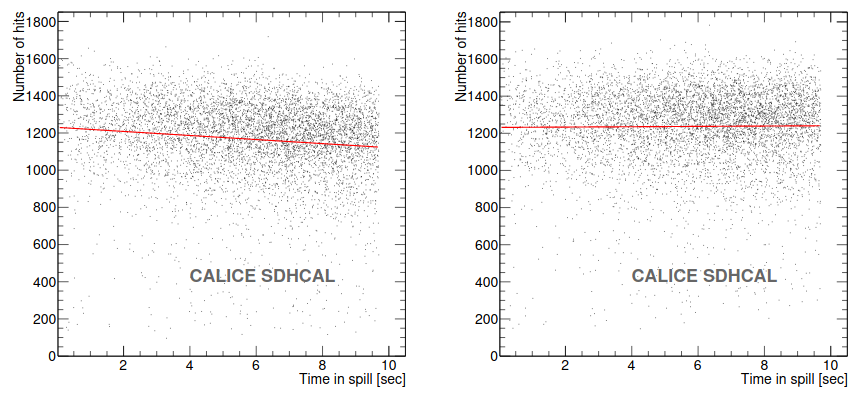}
    \caption{Total number of hits for 80 GeV pions as a function of the spill time before (left) and after (right) beam intensity correction.}
    \label{fig:SatCorrected}
\end{figure}

\noindent To correct this effect the average number of hits associated to each threshold as a function of the time $T$ in the spill is fitted to a straight line (Figure \ref{fig:Saturation}) and the slope of such fit $\lambda_j$ is computed for each threshold ($j = 1,\; 2\; \& \; 3$). Now, the number of hits per threshold is corrected using the following formula: $N_{corr_j} = N_j - \lambda_j * T$. The result is shown in Figure \ref{fig:SatCorrected} where the total number of hits in a 80 GeV pion is displayed before and after applying the correction. \\  

\subsection{Detector performance}

\begin{figure}
    \centering
    \includegraphics[width=0.75\linewidth]{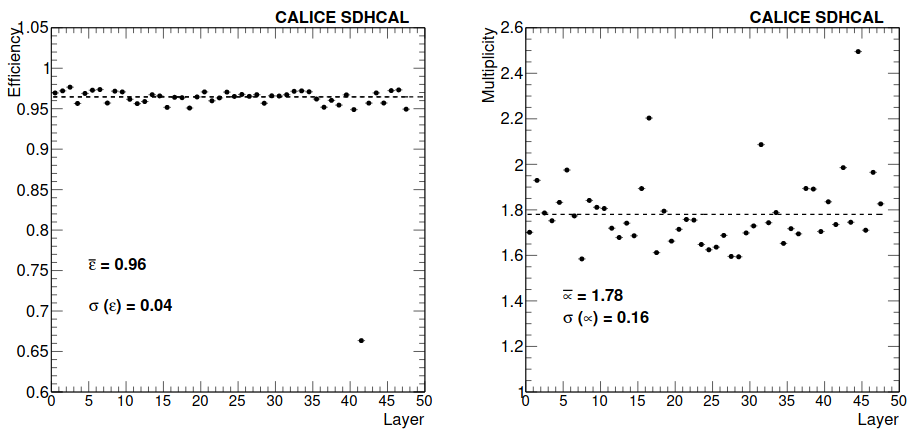}
    \caption{Mean efficiency (left) and average multiplicity (right) for each of the prototype's layers. The dashed line represents the mean value for all layers.}
    \label{fig:EffMult}
\end{figure}

To monitor the calorimeter performance, the efficiency ($\varepsilon$) and the hit multiplicity ($\mu$) of each layer are estimated using the beam muons \cite{2016}. For this study the muon tracks are reconstructed using the registered hits. First the hits are grouped into clusters if the cells share an edge and distant isolated clusters are dropped. Then, using all the selected clusters except the ones in the layers studied, the tracks are reconstructed and if there is a cluster at a distance of less than 3 cm in the studied layer then it is said to be efficient. The hit multiplicity for that layer is the size in number of hits from the cluster associated to the track. Figure \ref{fig:EffMult} shows the resulting mean value of efficiency and multiplicity per layer.

\subsection{Homogenization}

\begin{figure}[b]
\centering
    \includegraphics[width=0.4\linewidth]{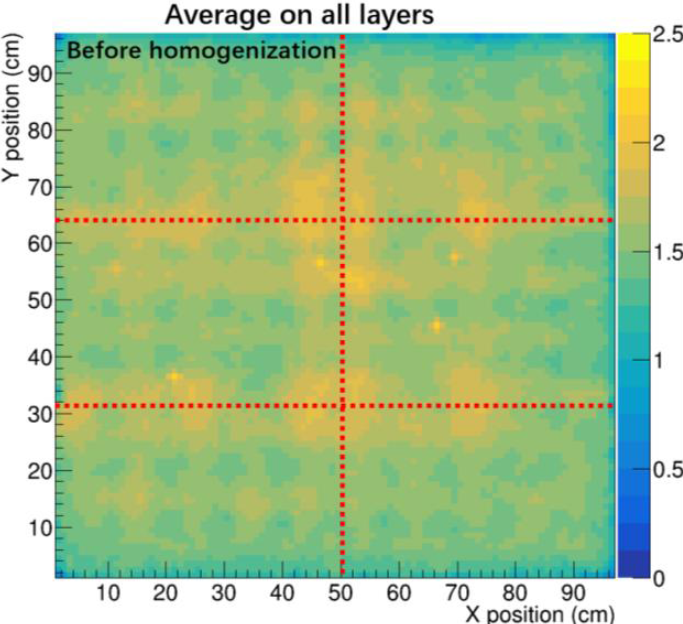}
    \includegraphics[width=0.4\linewidth]{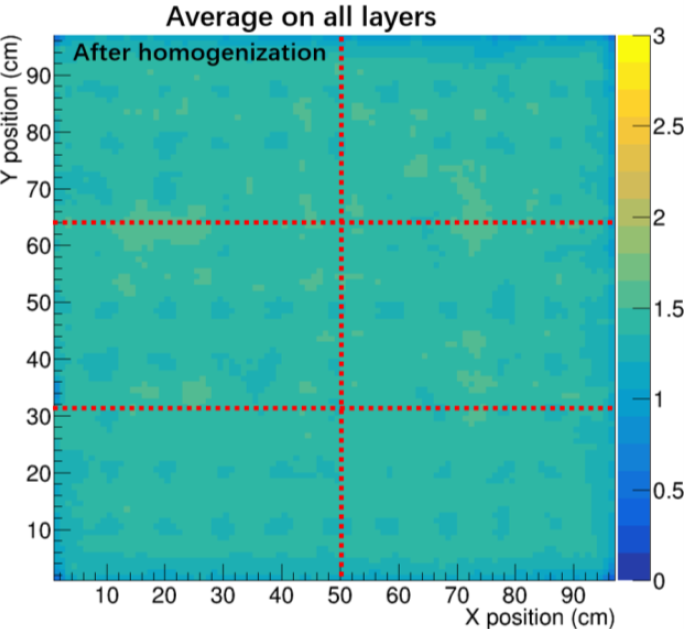}\\
    \includegraphics[width=0.4\linewidth]{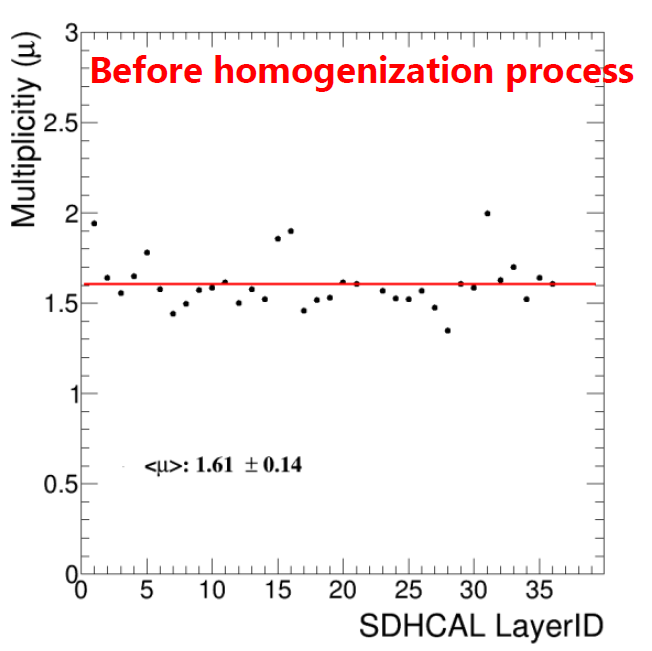}
    \includegraphics[width=0.4\linewidth]{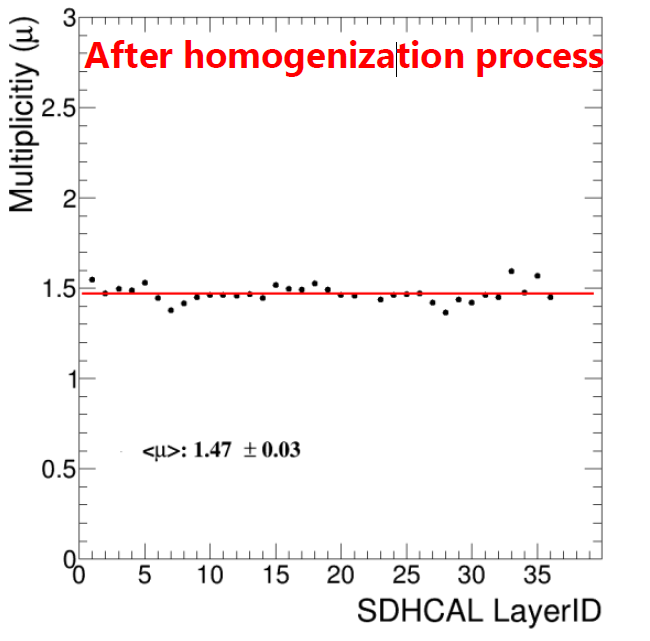}
    \caption{Average multiplicity for all layers as a function of the XY position before (left) and after (right) homogenization.}
    \label{fig:XYMult}
\end{figure}

The multiplicity computed in the previous analysis shows clear fluctuations from the mean value. This is due to different response from the detector depending on the layer and the position where the muon enters the layer as shown in Figure \ref{fig:XYMult} (left). Using the muons it is possible to study the multiplicity in the XY plane as a function of the threshold to find the set of values that homogenize the detector response, Figure \ref{fig:XYMult} (right).\\

\begin{figure}
    \centering
    \includegraphics[width=0.4\linewidth]{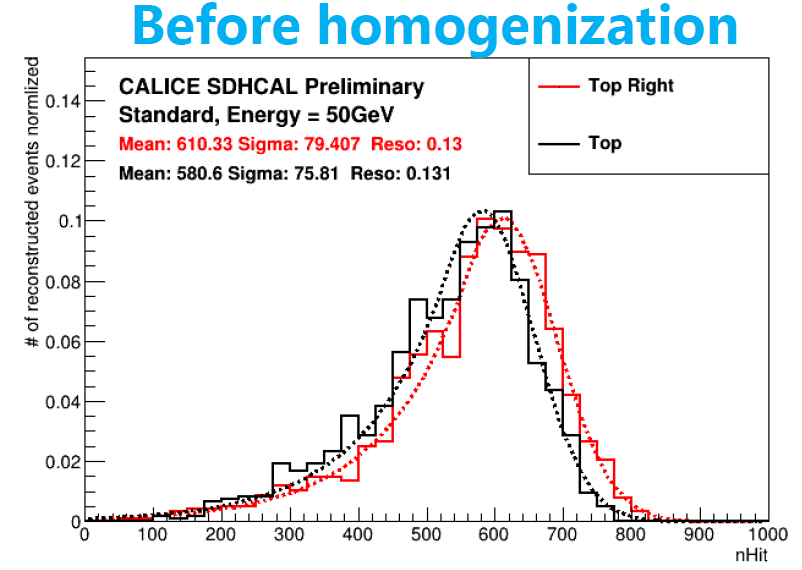}
    \includegraphics[width=0.4\linewidth]{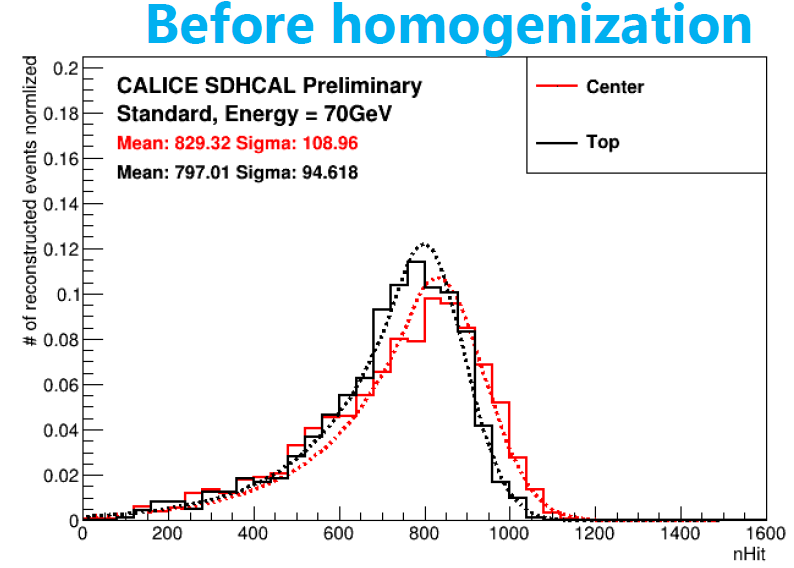}\\
    \includegraphics[width=0.4\linewidth]{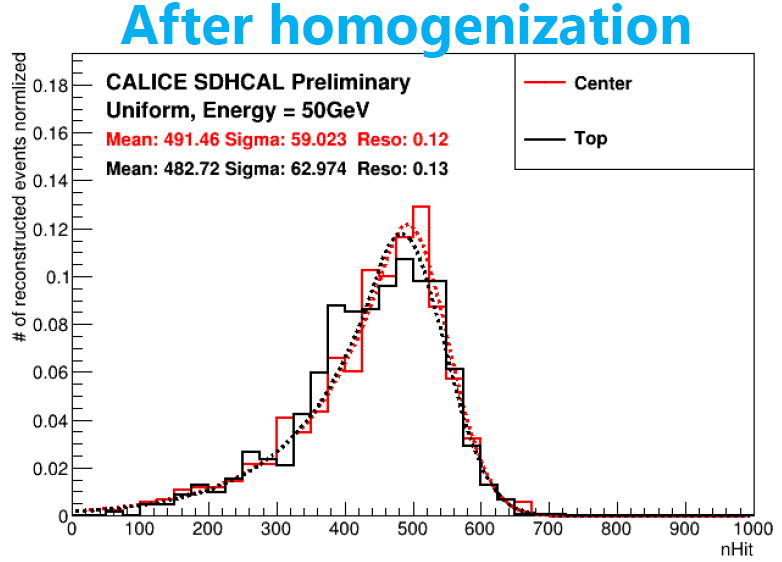}
    \includegraphics[width=0.4\linewidth]{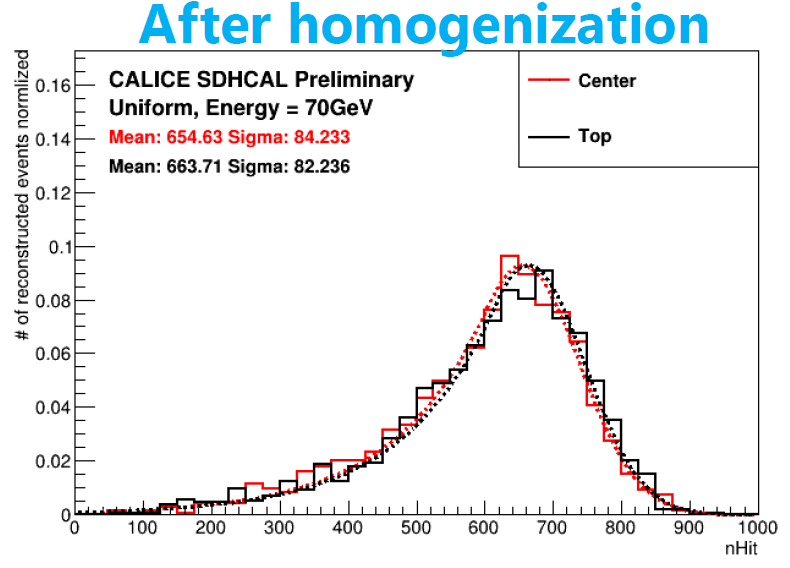}
    \caption{Total number of hits distributions for two runs comparing different incident beam positions before (top) and after (down) homogenization.}
    \label{fig:nhitHomogenization}
\end{figure}

\noindent This procedure equalizes the response of the detector in its transversal area. Figure \ref{fig:nhitHomogenization} shows that once taking it into account this effect does not longer affect the energy reconstruction from the number of hits distributions.

\section{Energy reconstruction}
\label{sec:EReco}

The particle selection presented in Section \ref{sec:ParticleIdentification} produces a set of hadronic showers in a wide range of energies (from a few GeV up to 80 GeV). To fully exploit the data provided by the SDHCAL the information related to the three thresholds can be used \cite{2016}. This information helps better estimate the total number of charged particles produced in a hadronic shower and improve the energy resolution. Pad crossed by a few particles are more likely to produces a hit in the second or the third threshold.\\

\noindent The reconstructed energy is parametrized as a weighted sum of the number of hits per threshold $N_1$, $N_2$ and $N_3$:

\begin{equation}
    E_{Reco} = \alpha \textcolor{green}{N_1} + \beta \textcolor{blue}{N_2} + \gamma \textcolor{red}{N_3}
\end{equation}

\noindent However, the complexity of the hadronic shower structure means that the optimal values of $\alpha$, $\beta$ and $\gamma$ are not constant over a large energy range, instead they are parametrized as function of the total number of hits $N_T = \textcolor{green}{N_1} + \textcolor{blue}{N_2} + \textcolor{red}{N_3}$:

\begin{align}
    \alpha &= \alpha_0 + \alpha_1 N_T + \alpha_2 N^2_T \nonumber \\
    \beta &= \beta_0 + \beta_1 N_T + \beta_2 N^2_T \\
    \gamma &= \gamma_0 + \gamma_1 N_T + \gamma_2 N^2_T \nonumber
\end{align}

\noindent To find the optimal values for this set of 9 parameters a $\chi^2$-like expression was used for the optimization method: 

\vspace{-10pt}
\begin{equation}
    \chi^2 = \sum^N_{i=1} \frac{(E^i_{Beam} - E^i_{Reco})^2}{E^i_{Beam}}
\end{equation}

\begin{figure}
    \centering
    \includegraphics[width=0.7\linewidth]{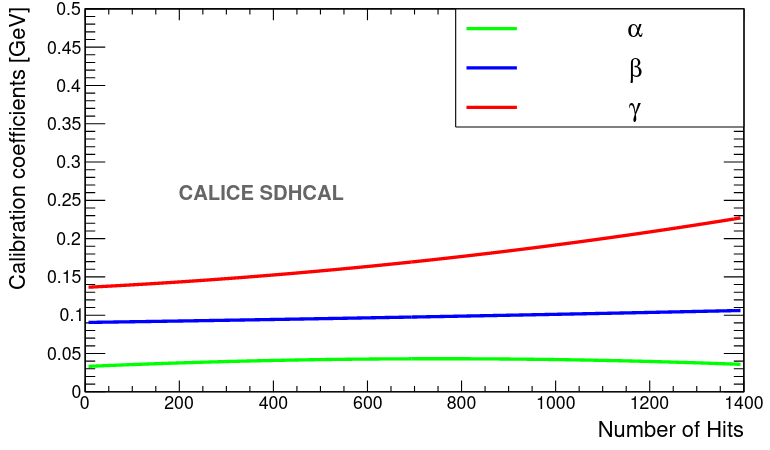}
    \caption{Polynomial functions computed from the minimization method associated to each threshold as functions of the total number of hits.}
    \label{fig:ERecoParameters}
\end{figure}

\vspace{10pt}
\noindent where N is the number of events used for the method. The procedure was applied to a third of the data available in one of the campaigns. The resulted values for the $\alpha$, $\beta$ and $\gamma$ functions are shown in Figure \ref{fig:ERecoParameters}. Once the set of optimal parameters are found it  is a straightforward process to transform the number of hits distributions into energy distributions. This energy distribution are fitted to a Crystall Ball function to take into account the tails produced by energy leakage.\\

\begin{figure}[b]
    \centering
    \includegraphics[width=0.4\linewidth]{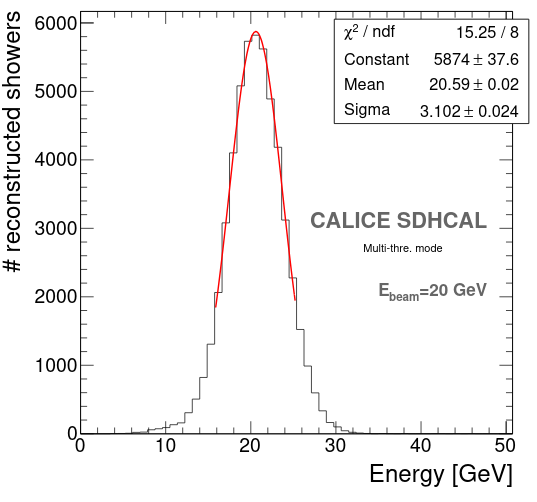}
    \includegraphics[width=0.4\linewidth]{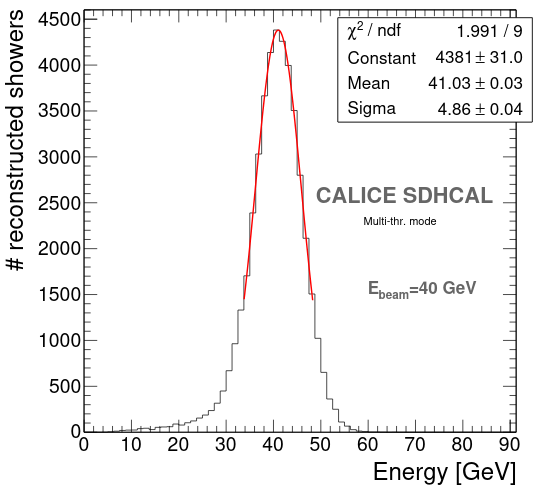}
    \caption{Energy distributions for a 20 GeV (left) and 40 GeV (right) pion runs. Only 1 $\sigma$ of the Gaussian part of the fits is shown.}
    \label{fig:ERecoCB}
\end{figure}

\begin{figure}
\centering
    \includegraphics[width=0.75\linewidth]{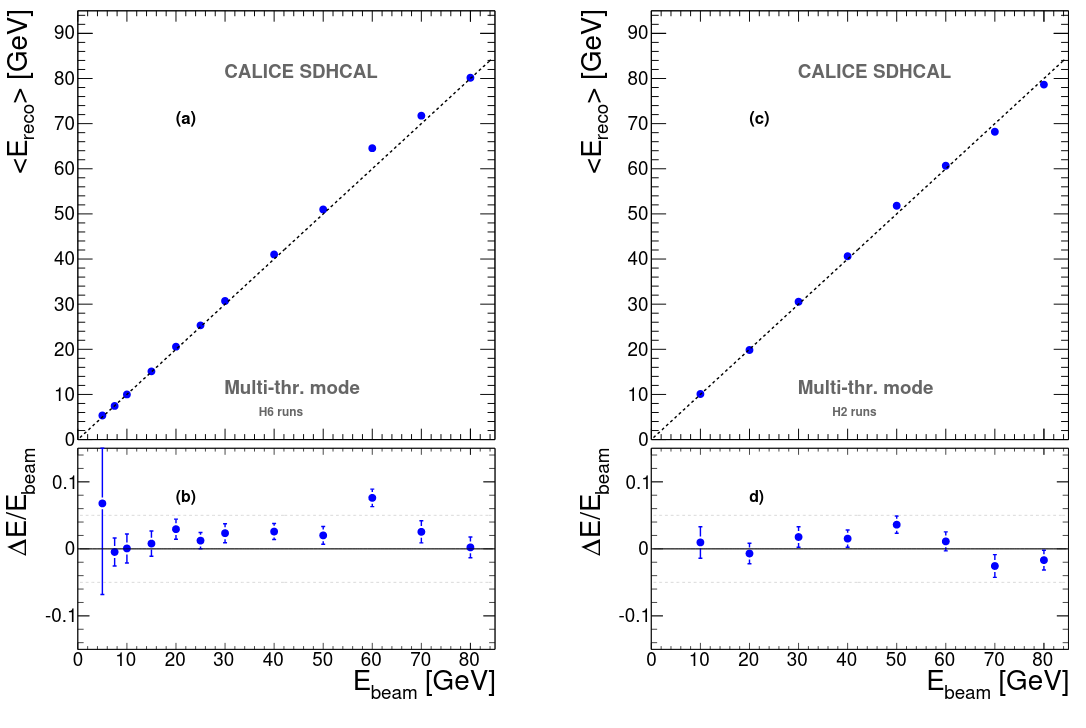}
    \caption{Linearity and deviations obtained in two data taking campaign. H6 runs made of positive pions (left) and the H2 campaign (right) of negative pion runs.}
    \label{fig:ERecoLinearity}
\end{figure}

\begin{figure}
\centering
    \includegraphics[width=0.75\linewidth]{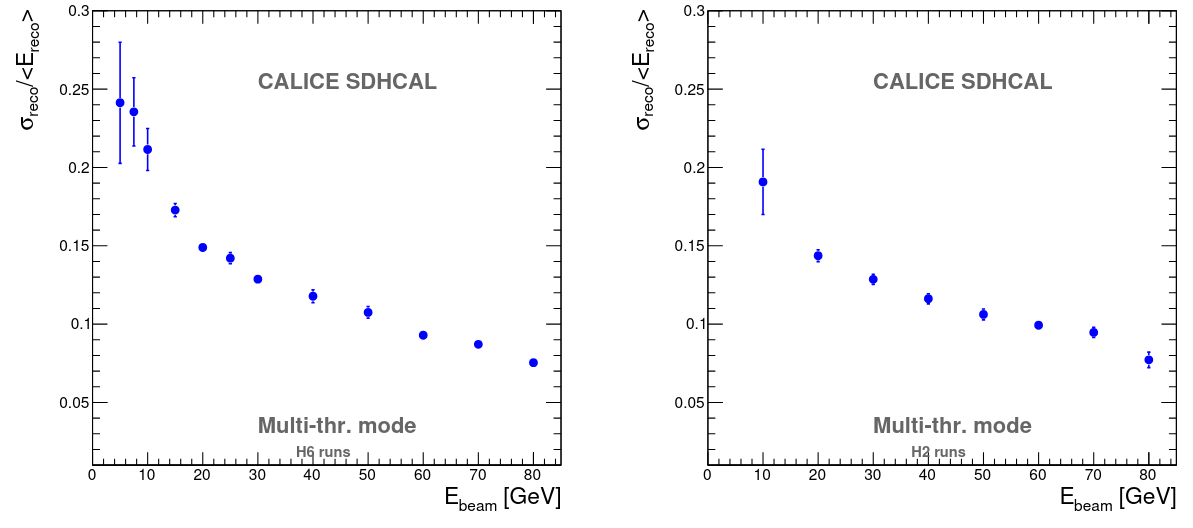}
    \caption{Resolutions obtained from the semi-digital parametrization for the two same campaigns as in the previous figure.}
    \label{fig:ERecoRes}
\end{figure}

\noindent This method of energy reconstruction of hadronic showers restores linearity over a wide energy range (from 5 GeV up to 80 GeV) as shown in Figure \ref{fig:ERecoLinearity} where it is shown deviation from linearity of less than 10\% in all the energy range. Also the usage of the three thresholds has a good impact on the energy resolution (Figure \ref{fig:ERecoRes}) at energies higher than 30 GeV where it reaches a value of 7.7\% at 80 GeV.

\section{Advanced methods}
\label{sec:Advanced}

This last section briefly describes the two most recent methods developed in the context of the SDHCAL: tracking using the Hough Transform Technique \cite{thecalicecollaboration2017tracking} and hadron selection using Boosted Decision Trees \cite{BDT}.

\subsection{Tracking using the Hough Transform Technique}

Hadronic showers often contain several track segments associated to charged particles. These particles cross several active layers before being stopped, react inelastically or escape the detector. However, even with the high granularity provided by the SDHCAL, it is difficult to separate the track segments from the highly dense environment present in hadronic showers. It was proposed to use the Hough Transform method to find tracks in noisy environments and here the results from that method are shown.\\

\begin{figure}[h]
    \centering  
    \includegraphics[width=0.4\linewidth]{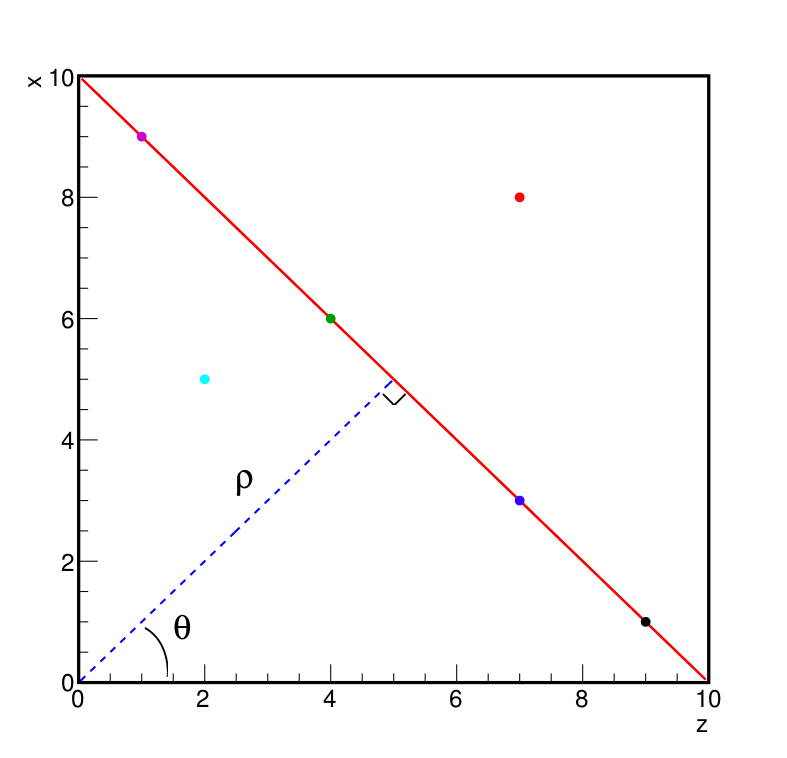}
    \includegraphics[width=0.4\linewidth]{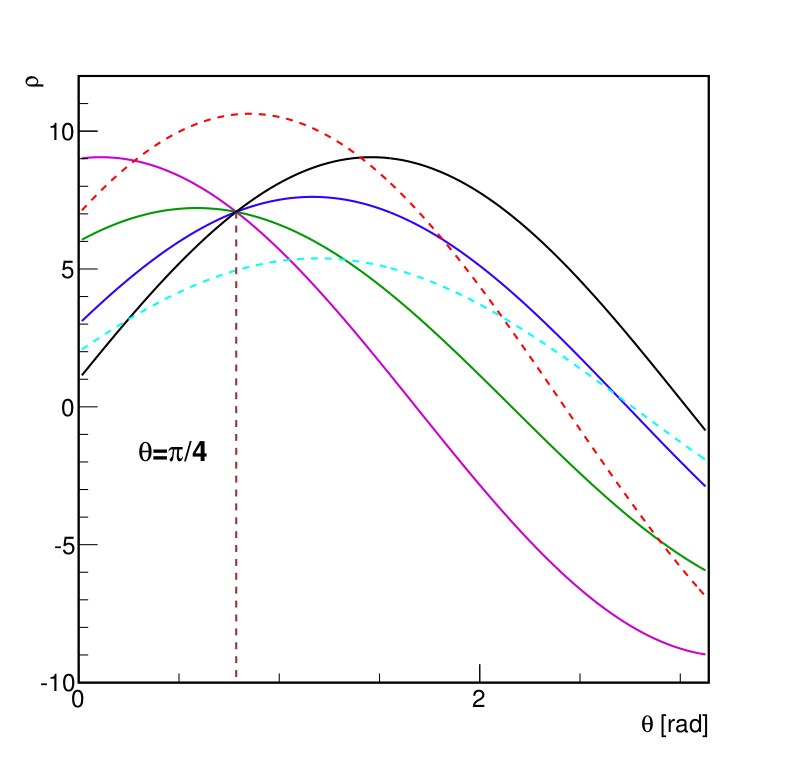}
    \caption{A point in the Cartesian plane (left) produces a curve in the polar plane (right).}
    \label{fig:HT}
\end{figure}

\begin{figure}
    \centering
    \includegraphics[width=0.4\linewidth]{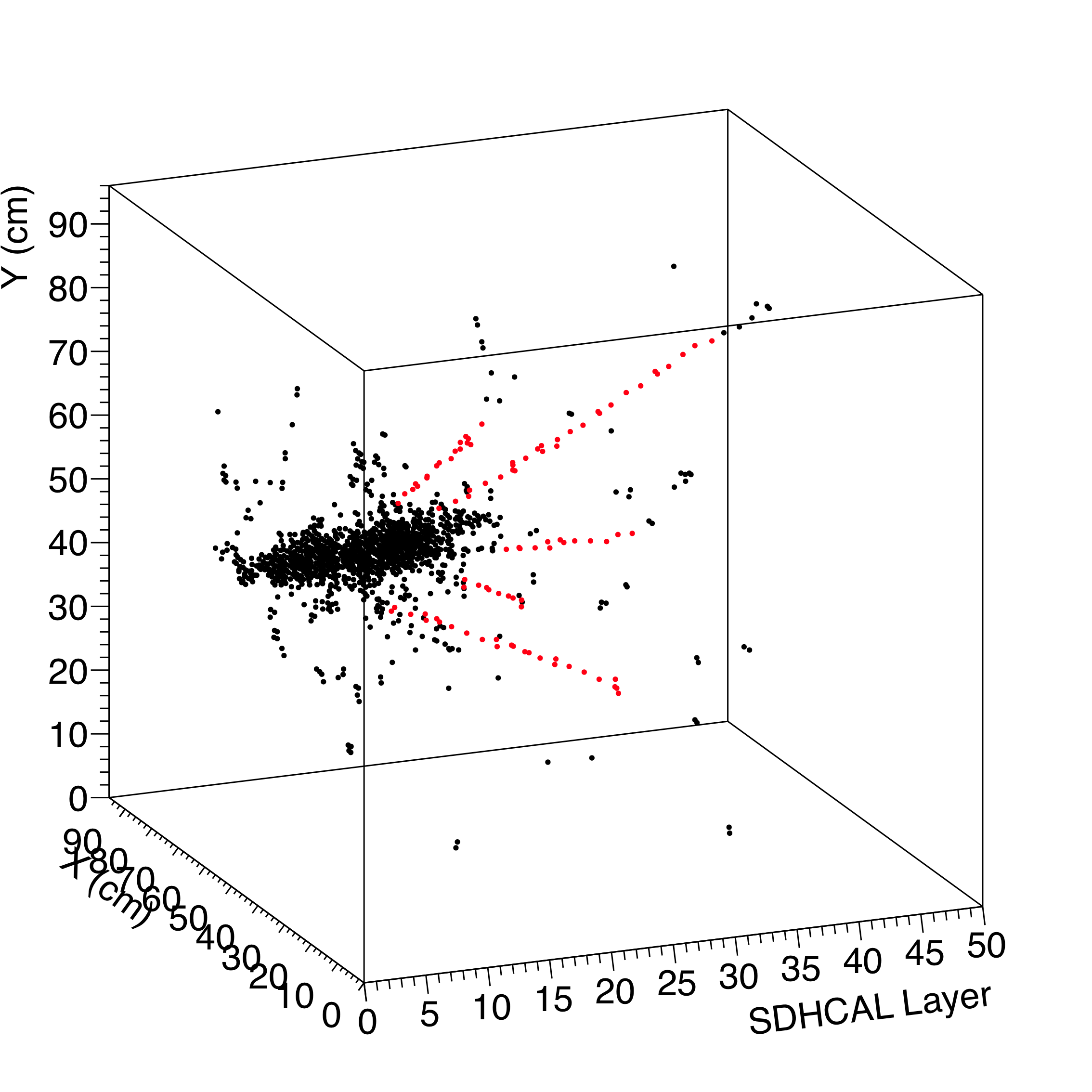}
    \caption{Tracks identified (red) in a 80 Gev pion shower.}
    \label{fig:HT_Display}
\end{figure}

\noindent Under a Hough transformation $\rho = zcos(\theta) + xsin(\theta)$ (Figure \ref{fig:HT}) a point in the Cartesian plane ($x$, $z$) defines a curve in the polar plane ($\rho$, $\theta$). Aligned points have their curves intersecting at one node ($\rho^0$, $\theta^0$) in the polar plane. This method needs to be adapted to a discrete set of position in a pixellated detector and, in the case of hadronic showers in the SDHCAL, it is only applied to cluster of hits outside the high density region of the shower to avoid creating artificial tracks. In Figure \ref{fig:HT_Display} the result of the process is shown for a 80 GeV pion shower where several tracks outside the core have been identified. \\

\begin{figure}
    \includegraphics[width=0.5\linewidth]{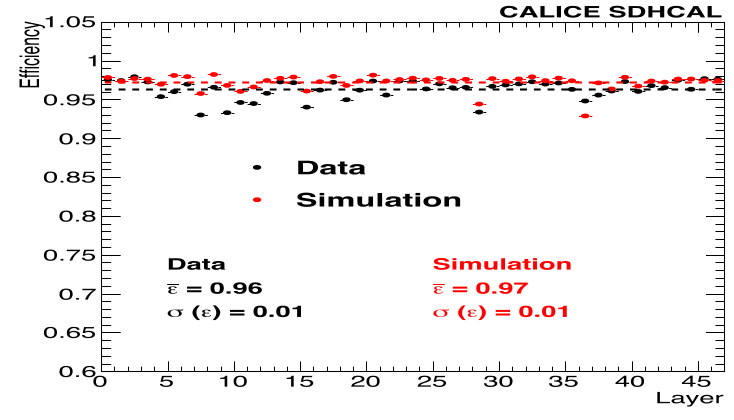}
    \includegraphics[width=0.5\linewidth]{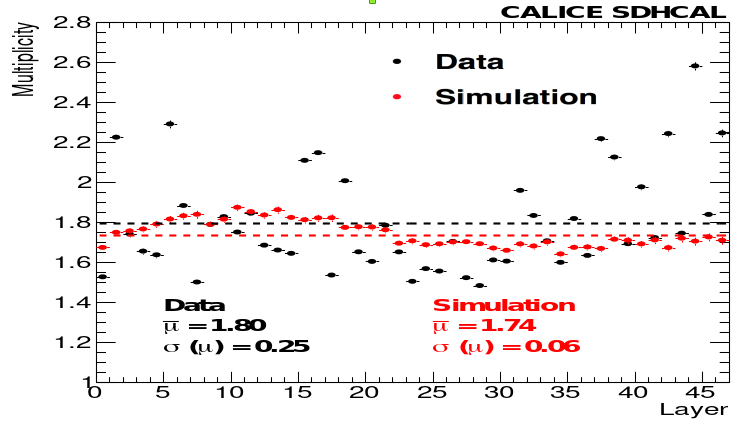}
    \caption{Efficiency and multiplicity mean values  per layer computed using track segments in a 40 GeV run compared with the simulation.}
    \label{fig:HT_Performance}
\end{figure}

\noindent The tracks extracted with this method have a multitude of uses. They play an important role in checking the active layer behaviour in situ by studying the efficiency and multiplicity of the detector. Average values of this two variables are shown in Figure \ref{fig:HT_Performance} per layer. The main difference of this method with the values computed in Section \ref{sec:CorrectionsPerformance} is that in this case the segments in the hadronic showers are not necessarily perpendicular to the active layer, producing slightly higher results.\\

\begin{figure}[b]
    \centering
    \includegraphics[width=\linewidth]{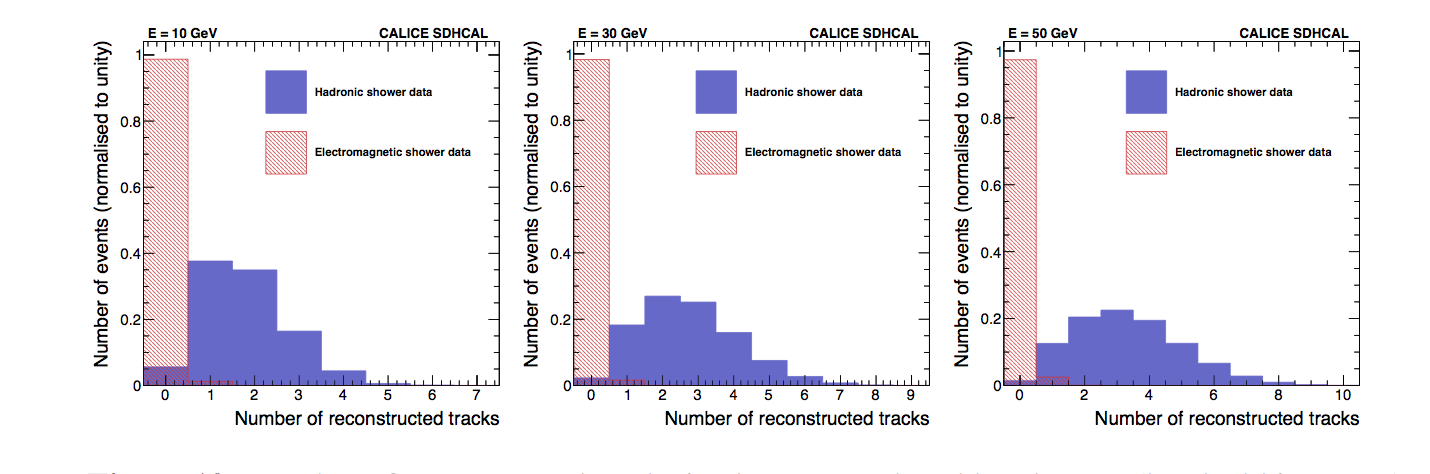}
    \caption{Number of reconstructed tracks in different energy electron and pion runs.}
    \label{fig:HT_NTracks}
\end{figure}

\begin{figure}[h]
    \includegraphics[width=0.5\linewidth]{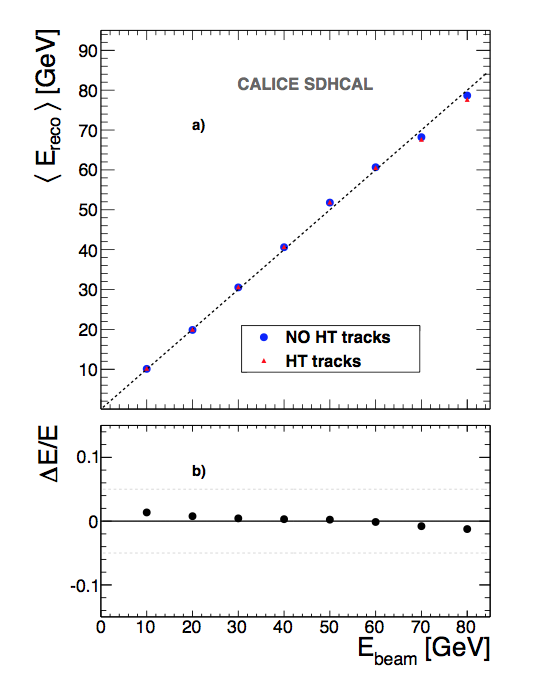}
    \includegraphics[width=0.5\linewidth]{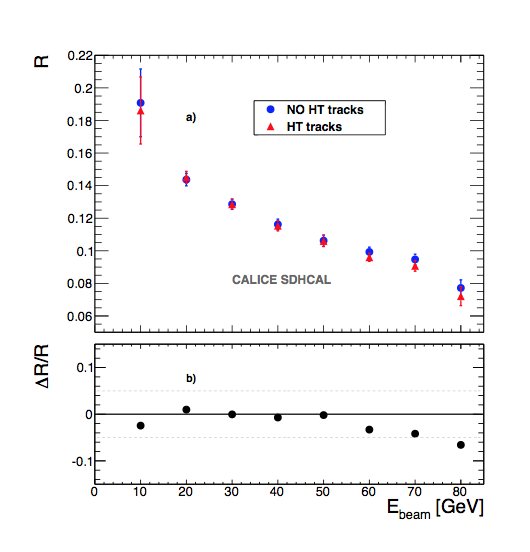}
    \caption{Linearity and resolution comparison between the standard method and the inclusion of track segments in the energy reconstruction.}
    \label{fig:HT_LinearityRes}
\end{figure}

\noindent Counting the amount of tracks segments reconstructed is a good tool to discriminate between electron and hadronic showers (Figure \ref{fig:HT_NTracks}). Also treating the hits in a track as belonging to a single particle independent of their threshold has an impact improving the energy resolution (Figure \ref{fig:HT_LinearityRes}).

\subsection{Hadron selection using Boosted Decision Trees}

Here an alternate method for the construction of the hadronic shower datasets based on Boosted Decision Trees is presented (BDT). Two training method were tested, one based on MC simulations and a data training approach. The variables used as inputs for the BDT to distinguish hadronic showers from electromagnetic showers and muons are the ones described below and Figure \ref{fig:BDT_Variables} compares them for different particle types to show their discrimination potential:\\

\begin{itemize}
    \item{\justify \textbf{First layer of the shower:} the first layer in the incoming particle direction containing at least 4 fired pads.}
    \item{\textbf{Number of tracks segments in the shower:} the number of tracks produced with the Hough Transform method explained in the previous subsection.}
    \item{\textbf{Ratio of shower layers over total fired layers:} a shower layer is a layer with a value of the root mean square greater than 5 cm.}
    \item{\justify \textbf{Shower density:} average number of neighbouring hits in a 3x3 grid including the hit itself.}
    \item{\textbf{Shower radius:} the root mean square of the hits with respect to the event axis.}
    \item{\textbf{Maximum shower position:} the distance between the shower start and the layer of the maximum radius of the shower.}
\end{itemize}

\begin{figure}[h]
    \centering
    \begin{tabular}{c c c}
        \includegraphics[width=0.3\linewidth]{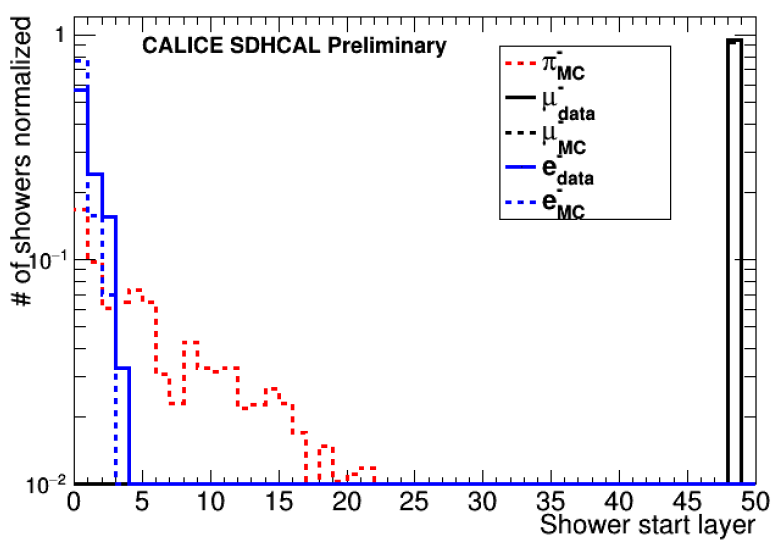} & \includegraphics[width=0.3\linewidth]{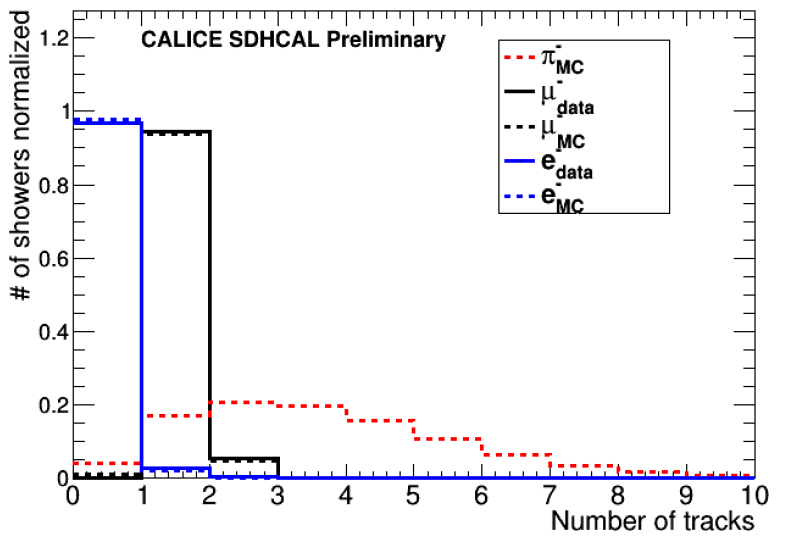} & \includegraphics[width=0.3\linewidth]{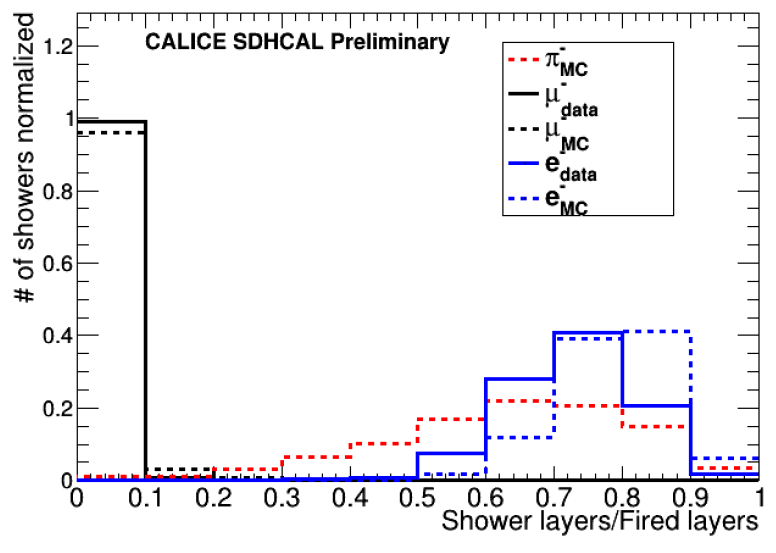} \\
        \includegraphics[width=0.3\linewidth]{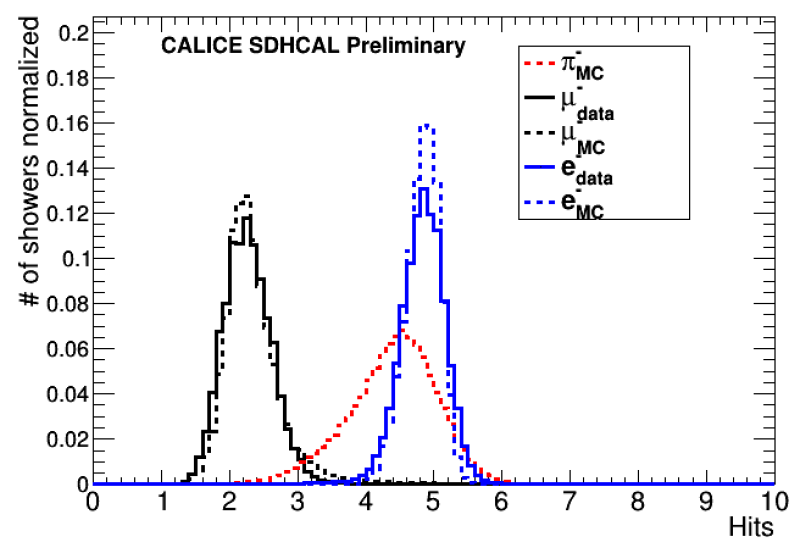} &
        \includegraphics[width=0.3\linewidth]{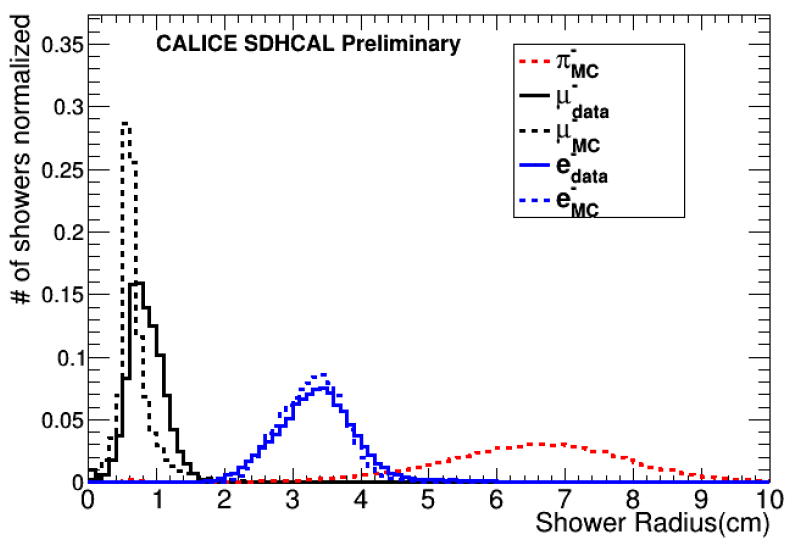} & 
        \includegraphics[width=0.3\linewidth]{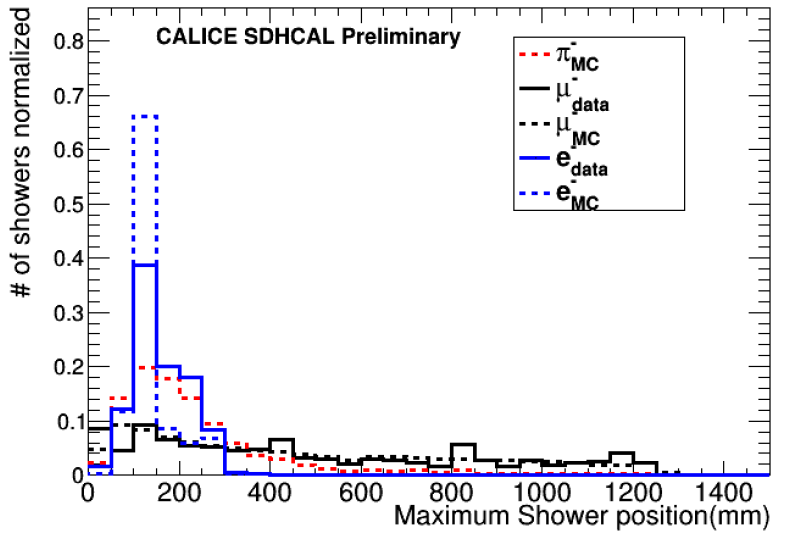}
    \end{tabular}
    \caption{Input variables for the BDT for each type of particle from data and MC.}
    \label{fig:BDT_Variable}
\end{figure}

\begin{figure}[b]
    \centering
    \includegraphics[width=0.75\linewidth]{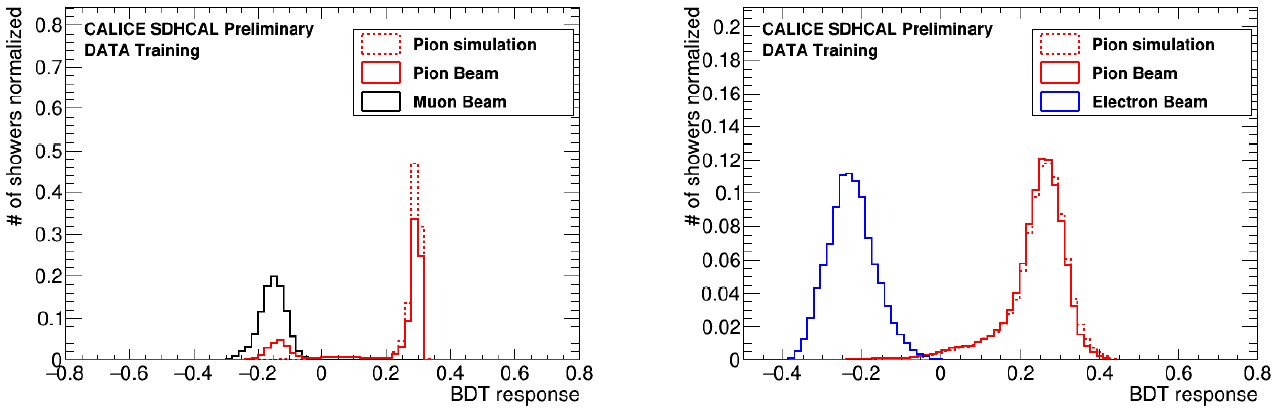}
    \caption{BDT output of the pions-muons sample (left) and of the pions-electrons one (right) in the data training method.}
    \label{fig:BDT_Output}
\end{figure}

\noindent Two classifiers are trained in both MC and data methods, one to separate the pions from electrons and another to separate pions from muons. The output of the classifiers is a variable in the range $[-1,1]$ where positive values represent pions while negative results identify the corresponding background, as shown in Figure \ref{fig:BDT_Output}. By applying a cut to the output variable the event is selected as signal or background, the following Table \ref{table:BDT_Cuts} summarizes this cuts:\\

\begin{table}[h]
    \centering
    \begin{tabular}{| c | c | c |}
    \hline
     & pion-muon & pion-electron \\
    \hline
    MC training & 0.1 & 0.05 \\
    \hline
    Data training & 0.2 & 0.05 \\
    \hline
    \end{tabular}
    \caption{Cuts applied to the BDT output to separate the signal from the corresponding background. The order of appliance is first the pion-muon and then the pion-electron classifier.}
    \label{table:BDT_Cuts}
\end{table}

\begin{figure}[h]
    \centering
    \includegraphics[width=0.75\linewidth]{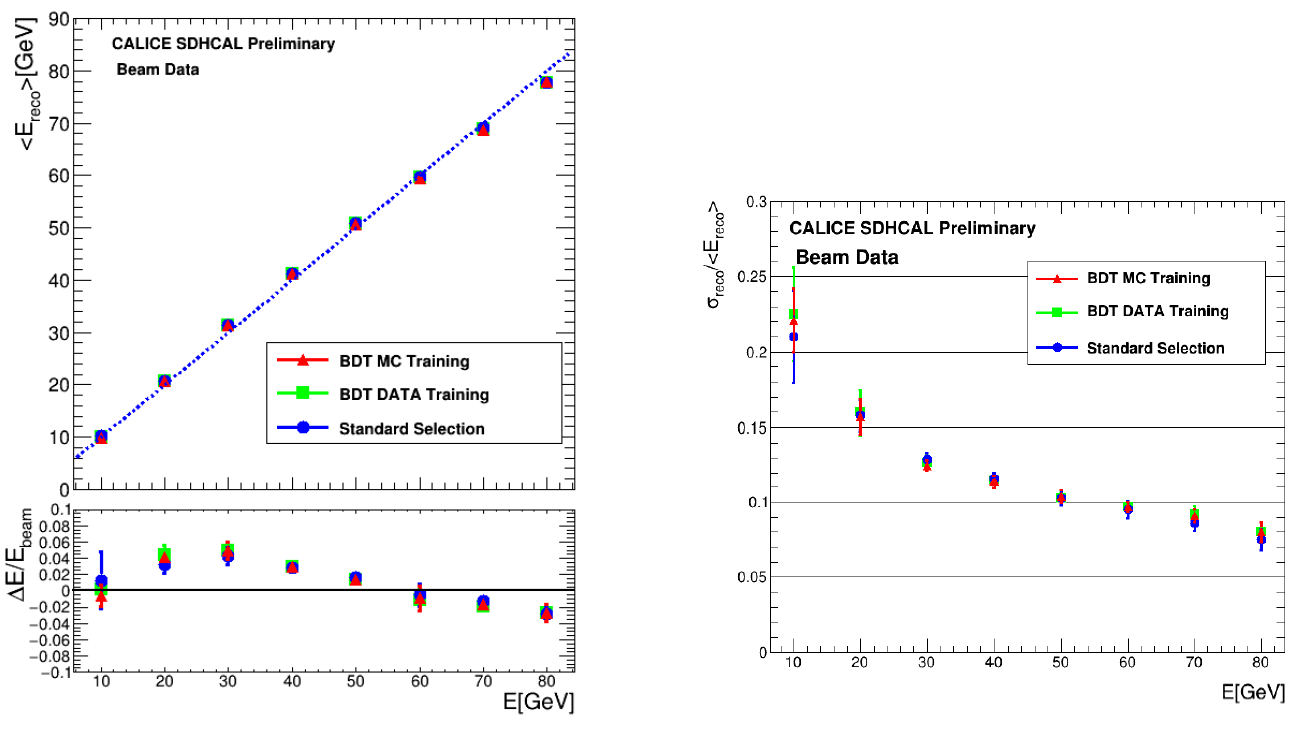}
    \caption{Comparison of linearity and resolution from the BDT MC training, BDT data training and the standard reconstructions.}
    \label{fig:BDT_Validation}
\end{figure}

\noindent Finally, to test this method the same energy reconstruction presented in Section \ref{sec:EReco} is applied (using the same parametrization). Similar results are obtained but the BDT method results have smaller statistical uncertainties (Figure \ref{fig:BDT_Validation}).

\section{Conclusion}

Different analysis, corrections and methods have been described, using data recorded with beams at CERN in several periods. Thanks to the high granularity of the detector, an analysis of its performance showed that the hit multiplicity behaved as expected and the muon analysis shows good values of the efficiency. A parametrization of the energy that restores the linearity for a wide range of energies have been found and produces a well behaved resolution reaching a value of 7.7\% at 80 GeV. Also the advanced methods developed have proved useful to improve the energy reconstruction, among other uses. However, currently a large chunk of the data is still being processed into new analysis and corrections that could potentially improve the energy resolution of the hadronic showers.

\newpage

\bibliographystyle{plain}
\bibliography{main.bib}

\end{document}